\documentclass[conference]{IEEEtran}
\IEEEoverridecommandlockouts
\usepackage{cite}
\usepackage{amsmath,amssymb,amsfonts}
\usepackage{algorithmic}
\usepackage{graphicx}
\usepackage{textcomp}
\usepackage{booktabs}
\usepackage{xcolor}
\def\BibTeX{{\rm B\kern-.05em{\sc i\kern-.025em b}\kern-.08em
    T\kern-.1667em\lower.7ex\hbox{E}\kern-.125emX}}
\begin{document}

\title{Context-Aware Systems for Sequential Item Recommendation\\
\thanks{All research was funded by Quizlet, Inc.}
}

\author{\IEEEauthorblockN{Moin Nadeem}
\IEEEauthorblockA{
\textit{Quizlet, Inc.}\\
San Fransisco, USA \\
moin.nadeem@quizlet.com}
\and
\IEEEauthorblockN{Dustin Stansbury}
\IEEEauthorblockA{
\textit{Quizlet, Inc.}\\
San Fransisco, USA \\
dustin@quizlet.com}
\and
\IEEEauthorblockN{Shane Mooney}
\IEEEauthorblockA{
\textit{Quizlet, Inc.}\\
San Fransisco, USA \\
shane@quizlet.com}}

\maketitle

\begin{abstract}
Quizlet is the most popular online learning tool in the United States, and is used by over $\frac{2}{3}$ of high school students, and $\frac{1}{2}$ of college students. With more than 95\% of Quizlet users reporting improved grades as a result, the platform has become the de-facto tool used in millions of classrooms.

In this paper, we explore the task of recommending suitable content for a student to study, given their prior interests, as well as what their peers are studying. We propose a novel approach, i.e. Neural Educational Recommendation Engine (NERE), to recommend educational content by leveraging student behaviors rather than ratings. We have found that this approach better captures social factors that are more aligned with learning.

NERE is based on a recurrent neural network that includes collaborative and content-based approaches for recommendation, and takes into account any particular student's speed, mastery, and experience to recommend the appropriate task. We train NERE by jointly learning the user embeddings and content embeddings, and attempt to predict the content embedding for the final timestamp. We also develop a confidence estimator for our neural network, which is a crucial requirement for productionizing this model.

We apply NERE to Quizlet's proprietary dataset, and present our results. We achieved an $R^2$ score of $0.81$ in the content embedding space, and a $recall$ score of $55\%$ on our 100 nearest neighbors. This vastly exceeds the $recall@100$ score of $12\%$ that a standard matrix-factorization approach provides. We conclude with a discussion on how NERE will be deployed, and position our work as one of the first educational recommender systems for the K-12 space.
\end{abstract}


\section{Introduction}
Founded in 2005, and used by more than $\frac{2}{3}$ of high school students, Quizlet, Inc. is the largest growing educational website in the United States\cite{HillaMeller}. The interactive platform permits students to learn any given "set", or collections of terms and definitions, in a variety of ways. However, with over 30 million monthly active users, and 250 million study sets, it has become nearly impossible for users to sift through all of the available content. This motivates a need for a system that will adapt to a user's preferences and make recommendations on what they should study next, given their prior history. 

This is not only motivated from a product perspective, but also by the rise of personalized learning. As a result of the rise of personalization in the e-commerce \cite{Linden2003}, social media \cite{Covington2016}, and dating \cite{Brozovsky2007}, many in education and research have grown curious about the implications personalized learning may have upon students.

Personalized learning can be defined as any functionality which enables a system to unique address each individual learner's needs and characteristics. This includes, but isn't limited to, prior knowledge, rate of learning, interests, and preferences. This provides the ability to ensure that each user's experience is best optimized for their unique needs and may save them time that would be otherwise wasted.

For an example that is applicable to Quizlet, one user may prefer to study content suitable to study with Spell Mode (where students practice spelling by typing the spoken word). Our algorithm would take that into account by biasing recommendations that are commonly studied in Spell Mode. Similarly, we may expect our algorithm to take user performance into account, and continue to recommend topics that the user hasn't quite mastered yet. 

The main contribution of this paper is a deep learning based system that provides personalized recommendations to Quizlet users, answering the question "What should I study next?". 

The rest of this paper is structured as follows: a summarization of previous literature for (educational) recommender systems is provided in Section 2. Section 3 provides an overview of our system architecture, model architecture, and dataset construction. We continue with a qualitative and quantitative assessment of our system in Section 4. Finally, we conclude our paper and provide a direction for future work in Section 5.
\section{Background}
Recommender Systems are a widely studied field, with contributions from major players such as Netflix \cite{Gomez-Uribe2015}, Google \cite{Covington2016}, and Amazon \cite{Linden2003}. The vast majority of these methods use matrix factorization techniques to decompose a user's preferences matrix, and an item ratings matrix into a latent space that represents how a user may rate a new item; this latent space is commonly derived from an Alternating Least Squares (ALS) algorithm. 

However, we believe that matrix factorization approaches aren't well suited for educational applications. To begin, the user-set matrix is extremely sparse. This makes standard matrix factorization based methods infeasible. These methods are also ill suited to material that is sequenced with temporal dependencies, as is usually the case for educational material. 

Instead, we attempt to make the problem computationally tractable by recurrent neural networks and set vectorization, which are able to learn both temporal dependencies and a dense representation of our data respectively. The rest of this section serves to summarize the current state of deep neural networks with respect to both the current state of recommender systems, as well as Technology Enabled Learning (TEL). We rely heavily upon previous contributions from the intersection of the two fields: Recommender Systems for Technology Enabled Learning (RecSysTEL).
\subsection{Literature Review}
Most recently, Tang \& Pardos \cite{Tang2017a} are the only other researchers in the RecSysTEL field who have explored the use of Recurrent Neural Networks (RNNs) for the purposes of personalization in learning. Their work leveraged RNNs to model navigational behaviors throughout Massively Open Online Courses (MOOCs). This research was conducted with the explicit intention of accelerating or decelerating learning as a result of performance in a given subject; the benefit to the user is a \textit{reduction in learning time and/or increased performance}. 

We believe that this work is quite notable due to the level of detail included in the model. Interactions as fine-grained as video pauses and changing video speed are included in the model as a proxy for mastery. However, Tang \& Pardos' algorithm was purely collaborative, and never leveraged the content of the MOOC(s) studied. We believe that this is an underexplored field in RecSysTEL, and aim for this to be a major contribution of our work.

Outside of the field of education, Covington, Adams, and Sargin \cite{Covington2016} at YouTube have developed the first recommendation system used in an industry setting that leverages deep neural networks. 

Covington \textit{et al}'s paper is interesting for two reasons. First, it demonstrates a successful use of a neural recommendation system at scale, thus mitigating any concerns about scaling such a system in production. Secondly, videos are quite analogous to Quizlet sets: both videos and sets represent ways to learn about topics, and may be episodic in nature.

To provide an example, if a user watched \textit{"Full House Episode 1"} on YouTube, a good recommendation would be \textit{"Full House Episode 2"}. Likewise, a good recommendation for a user who studied \textit{"Hamlet Chapter 1"} would be \textit{"Hamlet Chapter 2"}. In order to generate recommendations such as these, Covington et al. added search tokens as a feature to their network. 

In order to deal with the vast swaths of YouTube videos, Covington et al. split their network into two sub-networks. One network served to filter a large corpus of videos into those which the user may be interested in, and the second network (with access to many more features than the first) served to rank these candidates. Finally, their algorithm was both content-based and collaborative, demonstrating the viability of a hybrid approach. 

However, one major drawback of their method is the level of compute with which Google provides Covington, et al. This creates a challenge for us in creating a neural recommendation system while remaining within realistic computational resources.
\section{Methods}
In this section, we provide an overview of how we constructed our dataset, what our production system architecture will be, as well as how NERE is architected in detail.

\subsection{Dataset Construction}
In order to train NERE, Quizlet, Inc. assembled a proprietary dataset. Internally, we use Google BigQuery \cite{Sato2012} for all of our data warehousing needs. From BigQuery, we assembled two datasets from our activity logs: one which detailed our users and their respective metadata, and the second which detailed all sets studied by these users, and their respective metadata.

The users dataset contained the following fields:
\begin{table}[h!]
\centering
\begin{tabular}{@{}ll@{}}
\toprule
\textbf{Field}      & \textbf{Purpose}                                           \\ \midrule
User ID             & Uniquely mapping a row to a user.                          \\
Study Date          & Biasing the model to consistently recommend newer content. \\
Obfuscated IP Address          & Geo lookup to derive latitude, and longitude for locality. \\
Preferred Term Lang & Most common language to study terms in.                    \\
Preferred Def Lang  & Most common language to study definitions in.              \\
Preferred Platform  & Most common platform (Web, iOS, etc) to study on.          \\
Beginning Timestamp & Timestamp for when the study session started.              \\
Ending Timestamp    & Timestamp for when the study session ended.                \\
Set ID              & The set they studied during their session.                 \\
Session Length      & The number of minutes that their study session lasted.     \\ \bottomrule
\end{tabular}
\newline
\caption{Table \ref{users-dataset} contains information about all of our users and their metadata.}
\label{users-dataset}
\end{table}

The sets dataset contained the following fields:
\begin{table}[h!]
\centering
\begin{tabular}{@{}ll@{}}
\toprule
\textbf{Fields}           & \textbf{Purpose}                                                   \\ \midrule
Set ID                    & Uniquely mapping each set to a row.                                \\
Terms                     & All terms in a set as a space-delimited string.       \\
Definitions               & All definitions in a set as a space-delimited string. \\
Studier Count             & Number of unique users that have studied this set.                 \\
Broad Subject             & A high-level subject classification of the set.                    \\
Mean Studier Age          & The average age of the users who study the set.                    \\
Term Language             & The language that terms are in.                                    \\
Definition Langage        & The language that definitions are in.                              \\
Total Views               & The total number of views that this set has received.              \\
Has Images                & A boolean indicating whether this set contains images.             \\
Has Diagrams              & A boolean indicating whether this set contains diagrams.           \\
Preferred Study Mode      & The most common study mode used with this set.                     \\
Preferred Platform & The most common platform (Web, iOS, etc.) used.      \\
Mean Session Length       & The average session length for this set, in minutes.               \\ \bottomrule
\end{tabular}
\newline
\caption{Table \ref{sets-dataset} contains information about all of the sets and their metadata.}
\label{sets-dataset}
\end{table}

Once the datasets were assembled, we began cleaning the data. Since user privacy is quite important to Quizlet's values, we removed all users below the age of thirteen, and obfuscated Internet Protocol (IP) addresses by dropping the last octet. We believe that this is an important step towards preserving anonymity while still preserving quality recommendations.

All categorical variables, such as term language, were mapped to integers. All continuous variables were scaled between zero and one (with unit variance) to ensure smooth gradients. We replaced any missing continuous values with the mean of the dataset. Lastly, we mapped all IP addresses to their respective latitude and longitude, with the intuition that students in close proximity may be studying similar sets.

Finally, a preliminary test of NERE with this dataset found it difficult to model students who were studying for multiple classes on Quizlet. Intuitively, this makes sense, as the recurrent neural network is looking for temporal relations in places where these relations were murky at best. We solve this by separating sequences by their \texttt{broad subject}\footnote{The \texttt{broad subject} field was of the following enumerated type: Theology, History, Uncommon Languages, Communications, Formal sciences, Visual Arts, Social Sciences, Applied Sciences, Vocabulary, German, Performing Arts, Sports, French, Reading Vocabulary, Spanish, Natural Sciences, and Geography.} column. This was done in practice by concatenating each User ID with the subject they studied, ensuring each row is unique in both user and subject classification. After cleaning, we were left with 1,616,004 unique user-subject combinations to be fed into our model.

To vectorize our Words and Definitions, we took the space-delimited string and removed stopwords and non-ASCII characters. Next, we tokenized it and trained 128-dimensional GloVe embeddings, which effectively creates an implementation of \textit{Set2Vec}. These embeddings were concatenated along with the preprocessed set metadata to create our set vectors.

Finally, we transformed our dataset into a timeseries format by concatenating all user study sessions into a single axis and sorting by ending timestamp. We chose a session length of 5 timesteps, since 90\% of our users have at least five sessions. The dimensions of the resultant datasets are as follows:
\begin{itemize}
	\item User Metadata: (1616004, 5, 13)
	\item Set Metadata: (1616004, 5, 12)
	\item Set Content Vectors: (1616004, 5, 128)
\end{itemize}

\subsection{System Architecture}
For deployment purposes, we have the following system architecture.

\begin{figure}[h!]
\centering
\includegraphics[width=0.5\textwidth]{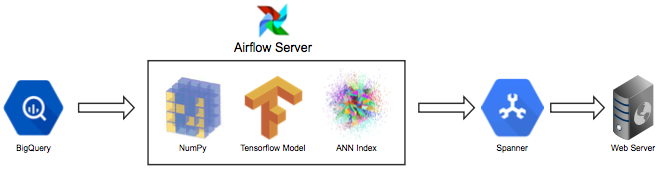}
\caption{This figure depicts how our model is used to serve recommendations in production.}
\label{production-environment}
\end{figure}

Quizlet uses Apache Airflow \cite{Takamori2016}, the industry standard for Extract-Transform-Load (ETL) pipelines, to schedule jobs. Every week, Apache Airflow reads datasets from BigQuery. Within Airflow, this dataset is preprocessed, and sent to TensorFlow. TensorFlow predicts which sets the user should study next, and sends the embedding back to Airflow. Airflow maps the vectors to sets by determining the $N$ nearest neighbors of this embedding, and subsequently caches these recommendations to spanner. Finally, our web server reads these recommendations from Spanner when serving content. Figure \ref{production-environment} depicts this flow visually.

Our web server reads from this cache when serving user content. Since the model takes \textit{2ms} to predict on each user with a CPU, we have opted to use a CPU-backed instance rather than a GPU-backed instance due to infrastructure cost.

\subsection{Algorithm Architecture}
In this subsection, we first introduce a formalization of our set-based recommendation task. Then, we describe our proposed NERE model architecture in detail.

\subsubsection{Formalization}

Session-based recommendation is the task of predicting what a user would like to study next when their previous history and metadata are provided. 

We let $X$ = [$s_1, s_2, s_3, ..., s_{n-1}, s_n$] be a study session, where $s_i \in S$ ($1 \leq i \leq n$), $n$ is the input length, and $S$ represents the pool of study sessions. We learn a function $f^{\hat W}(\cdot)$ such that for any given set of $n$ prefixes, we get an output $Y = f^{\hat W}(X)$.

Since our recommender will need to predict several states $[s_{n+1}^0, s_{n+1}^1, ..., s_{n+1}^m]$ for the $(n+1)^{th}$ timestep, where $m$ is the number of recommendations desired, we must be able to derive several Quizlet sets from $Y$. We let $Y$ be a 128-dimensional vector that represents the content for a Quizlet set and perform NNDescent \cite{Dong2011} for a fast, approximate $m$-nearest neighbors search algorithm on $Y$. We find that this provides an efficient manner to recommend multiple sets while maintaining a dense representation for the model to learn.
\subsection{Model Architecture}
\begin{figure}[h!]
\centering
\includegraphics[width=0.3\textwidth]{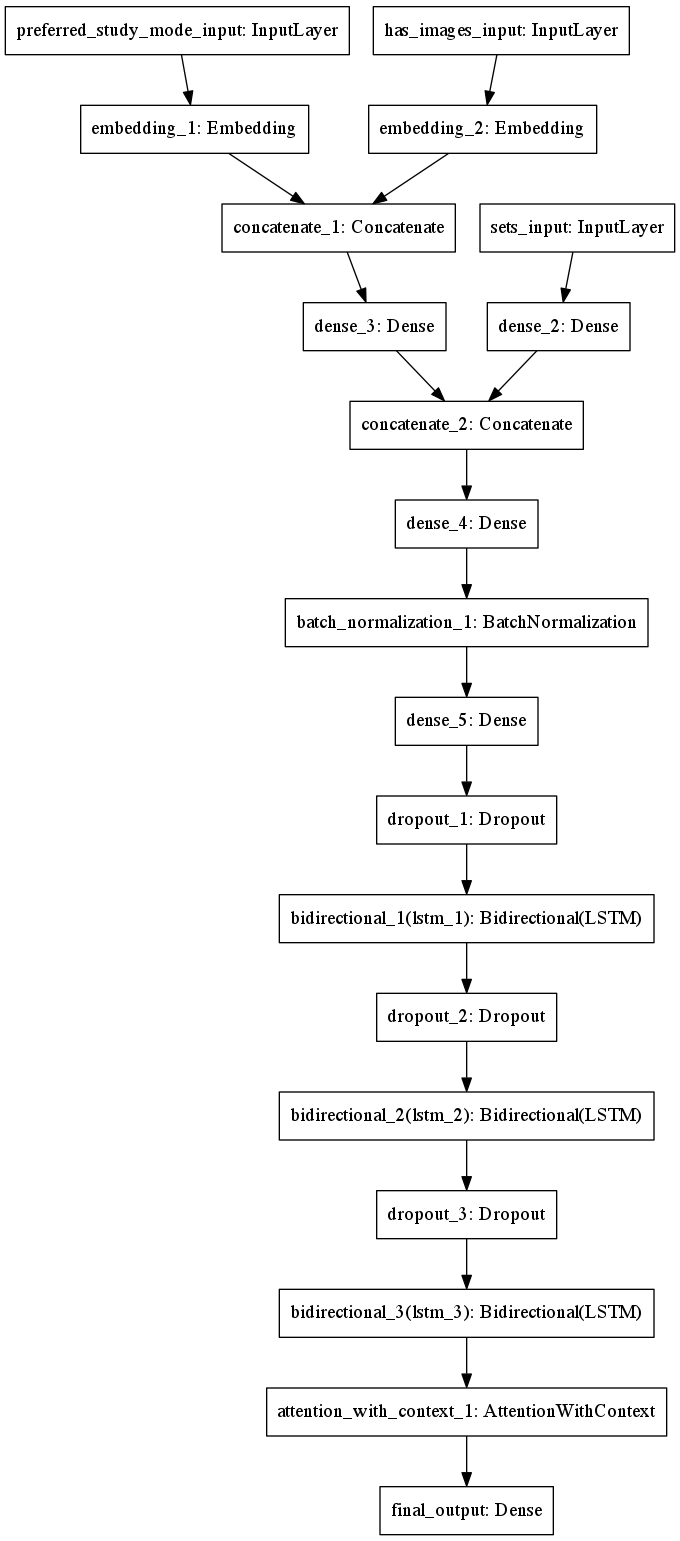}
\caption{This figure provides a slice of our model architecture; some inputs have been excluded for brevity.}
\label{model}
\end{figure}

Our model consists of 56 layers, 22 of which are inputs to the model. Figure \ref{model} depicts a portion of our model architecture.

In our architecture, we employ quite a few non-standard layers popular in Natural Language Processing. The remainder of this subsection will be explaining these layers.
\subsubsection{Embedding Layer}
In order to provide a dense representation for our categorical variables, we trained a embedding matrix \cite{Lopez-Sanchez2017}. 

Each categorical variable $C_i$ $\in$ $C$, where $C$ is the set of categorical variables, was mapped to a 32-dimensional representation. This was done with the explicit intention that the model may learn a spatial relation for some of these variables.

Each category $c_j \in C_i$ ($1 \leq j \leq |C_i|$) is learned using the following table: 
\begin{equation}
LT_{W^i}(j) = W_{j}^i
\end{equation}

Where $W^i \in \mathbb{R}^{32 \times |C_i|}$, $|C_i|$ represents the number of categories in $C_i$, and $W_j^i$ is the $j^{th}$ column of matrix $W^i$ that represents the 32-dimensional vector corresponding to category $c_j$. It is important to note that the entirety of this matrix is randomly initialized, and the vectors are learned jointly through backpropagation. 
\subsubsection{Bidirectional Layers} Bidirectional Layers \cite{Schuster1997} are commonly utilized to help models learn sequences.

The intuition behind bidirectional layers is that it helps recurrent layers learn sequences by making the context more explicit. It splits a recurrent layer into a part that is responsible for the learning the input normally, and another part that is responsible for learning the input backwards; this helps the model understand what \textit{may} happen in the future. 

Formally, given some study sequence $x_1, x_2, x_3, ..., x_{n-1}, x_n$, it would feed [($x_1$, $x_n$), ($x_2$, $x_{n-1}$), ..., ($x_n$, $x_1$)] as the input. At first sight, one would believe that this leaks information; however, humans do precisely the same by inferring future states from previous experience.

\subsubsection{Attention With Context} Based off of the work of Yang, \textit{et al.}, Attention With Context is a mechanism that helps the model learn which features are important, and which ones may be discarded. As the name may imply, it helps the model \textit{pay attention}.

Formally, we add a new layer that performs the following operation. We assume that $i$ is the $i^{th}$ timestamp in our input, and $t$ is the $t^{th}$ element in the vector $i$. Lastly, $h_{it}$ is the output of the $i^{th}$ element of the $t^{th}$ timestamp in the layer that precedes our attention layer. The following equations describe the operations of the Attention layer:

\begin{equation}
u_{it} = tanh(W_w h_it + b_w)
\end{equation}
\begin{equation}
\alpha_it = \frac{exp(u_i u_w)}{\sum_t exp(u_it u_w)}
\end{equation}
\begin{equation}
s_i = \sum_i{\alpha_{it} h_{it}}
\end{equation}

Where $u_w$ is a learned feature-level attention vector, $W_w$ are the weights of the attention layer, and $\alpha_{it}$ is a weighted $t^{th}$ element of the $i^{th}$ vector. Intuitively, this implementation makes a lot of sense: the model is computing how important each feature in each timestep is against all other features in the same timestep, and re-weighing the input accordingly. All weights in this layer are randomly initialized and jointly learned throughout the training process.

\subsubsection{Miscellaneous Features}
While most other works have used Long-Short Term Memory (LSTM) \cite{Hochreiter1997} cells for their recurrent unit, we chose to use Gated Recurrent Unit (GRU) \cite{Cho2014} cells. As Chung \textit{et al.} show in \cite{Chung2014}, for short sequences, GRU cells commonly are more practical due to not having an internal \textit{memory}. We saw a noticeable speed up of more than 20\% when using a GRU cell over an LSTM.

In order for these models (over 5,994,444 learnable parameters) to generalize, we had to apply some strict regularization. We applied 50\% dropout on layers following a recurrent cell, and applied 0.001 L2 regularization on the recurrent kernel itself. Furthermore, we used batch normalization to ensure that our inputs are zero-centered with normalized variance. Following the results of Santurkar \textit{et al.} \cite{Santurkar2018}, we also noticed faster training times as a result of these smoother gradients.

\section{Results}
In this section, we evaluate NERE from a qualitative and quantitative perspective. We compare our model against a baseline matrix factorization approach, and analyze several variations of the model for the purposes of introspection.

Table \ref{recommendations} shows the qualitative results of our recommendation system. The \textbf{studied} column shows the set that the user studied, while the \textbf{recommendation} column shows the set that was recommended for the user to study. For this particular recommendation, our system understands that a student had been learning about discussing time (in terms of days of the week) in French, and recommended a corresponding set about months of the year. This shows that the model understands that the user is learning about temporal relations. On a higher level, this demonstrates a level of understanding of both the content that a user desires to learn and the difficulty at which he desires to learn it.

\begin{table}[h!]
\centering
\begin{tabular}{@{}clll@{}}
\toprule
\multicolumn{4}{c}{\textbf{Recommendation Results}} \\ \midrule
\multicolumn{2}{c}{\textbf{Studied}} & \multicolumn{2}{c}{\textbf{Recommendation}} \\
\multicolumn{1}{l}{Term} & Definition & Term & Definition \\ \bottomrule
lundi	&	Monday	&	au printemps	&	spring	\\
mardi	&	Tuesday	&	en été	&	summer	\\
mercredi	&	Wednesday	&	Les mois	&	the months	\\
jeudi	&	Thursday	&	Janvier	&	January	\\
vendredi	&	Friday	&	Février	&	Febuary	\\
samedi	&	Saturday	&	Mars	&	March	\\
dimanche	&	Sunday	&	Avril	&	April	\\
un an	&	a year	&	Mai	&	May	\\
une année	&	a year	&	Juin	&	June	\\
après	&	after	&	Juillet	&	July	\\
avant	&	before	&	Août	&	August	\\
après-demain	&	the day after tomorrow	&	Septembre	&	September	\\
un après-midi	&	an afternoon	&	Octobre	&	October	\\
aujourd'hui	&	today	&	Novembre	&	November	\\
demain	&	tomorrow	&	Décembre	&	December	\\
demain matin	&	tomorrow morning	&	Quand	&	When	\\
demain après-midi	&	tomorrow afternoon	&	Où	&	Where	\\
demain soir	&	tomorrow night	&	Comment	&	How	\\
hier	&	yesterday	&	Avec qui	&	With whom	\\
\bottomrule
\end{tabular}
\newline
\caption{Table \ref{recommendations} shows the results of our recommendation system.}
\label{recommendations}
\end{table}

We use two proxies to assess model accuracy: \textit{recall@100} and $R^2$. In order to compute \textit{recall@100}, we take the 100 nearest neighbors of our output embedding, and check if the set that the learner studied at timestep $T_{n+1}$ is in the set of nearest 100 neighbors. If it is, we mark that recommendation as correct; otherwise, it is incorrect. We use the 100 nearest neighbors due to the density of our embedding space, as well as the fact that many of the sets in our embedding space are near-duplicates due to a lack of canonicalization.

We use $R^2$ to assess whether the predictions in the embedding space match the actual distribution; this serves as a sanity check to ensure that our model's output distribution is correlated to the expected distribution.

\subsubsection{Comparison Against Matrix Factorization}
We compare the performance of NERE against that of TensorRec \cite{Kirk2017}, a library written by James Kirk that uses the Tensorflow API. TensorRec accepts a user matrix, item matrix, and interactions matrix as inputs, and formulates a predictions matrix as an output. For the user matrix, we provide the user metadata matrix that NERE is provided. We concatenate the set vectors and set metadata, and this represents the item matrix. Lastly, we create an interactions matrix of dimensions ($|USERS|$, $|SETS|$), where some $(i,j)=1$ if user $i$ studied set $j$. 

We trained TensorRec on this dataset, and it obtained a \textit{Recall@100} of 0.12 after convergence. We believe this validates our belief in a core difference between a matrix factorization approach and our approach: even after extensive customization, an approach based off of temporal data is much more likely to provide quality recommendations for educational content.
\subsubsection{Input Sequence Length}

Our NERE model is based off of the assumption that a user is purposefully selecting sets to study, and  topically related to a greater theme. This permits us to also believe that the sets are temporally related, and therefore, enables us to use a recurrent neural network. 

\begin{figure}[]
\centering
\includegraphics[width=0.5\textwidth]{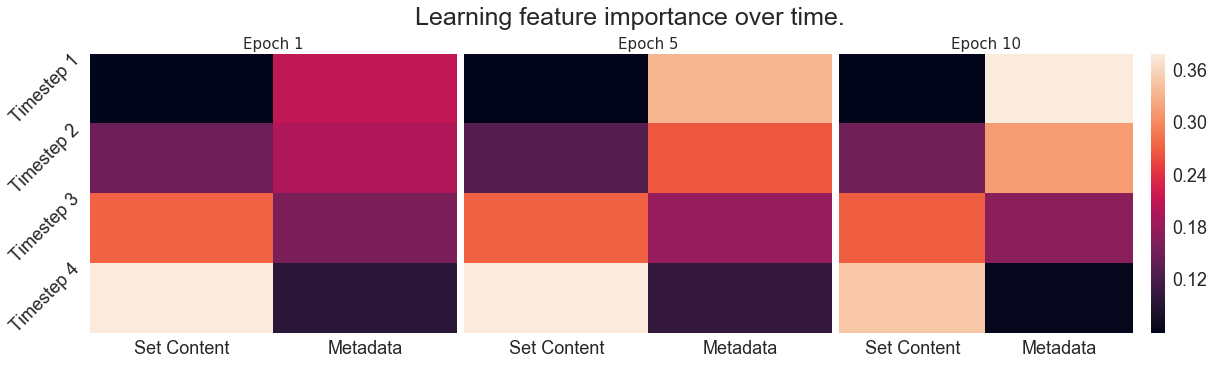}
\caption{This figure visualizes the model's internal attention vector.}
\label{performance}
\end{figure}

Figure \ref{performance} validates this assumption by comparing model performance against the input sequence length. We see that the $R^2$ score slowly converges, but that the \textit{recall@100} metric steadily increases until our fourth input sequence. This implies that there may be performance advantages to be obtained by increasing the length of the input sequence past four. However, since we begin to lose a significant number of users in our dataset if we extend beyond five timesteps, we risk creating a model that will not generalize to our entire userbase. As a result, we believe that five timesteps is a good balance between desired accuracy and generalizability.

\begin{figure}[]
\centering
\includegraphics[width=0.5\textwidth]{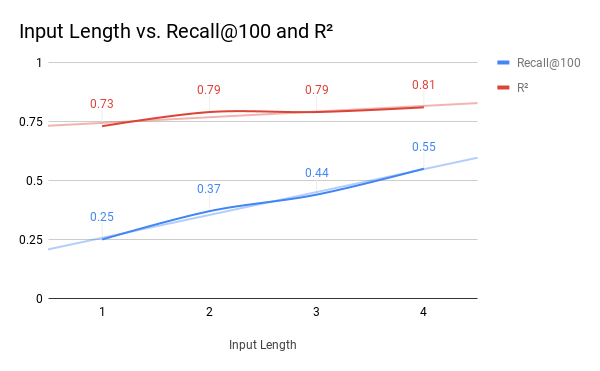}
\caption{This figure visualizes how the length of the input may affect model performance.}
\label{attention}
\end{figure}

\subsubsection{Where's the Attention}
One popular use of attention in deep neural networks is to visualize the model's understanding of the input. 
Figure \ref{attention} visualizes how the model pays attention to the input, as well as how it learns the attention vector over time. Brighter rectangles indicate that more attention is being placed on those blocks. 

These results show incredible insight into the decision process of the model. We can see that at the beginning of the input, the model focuses on the metadata; aspects such as term and definition language are deemed incredibly important. However, as time goes on, the attention shifts from set and user metadata towards content-based features. We see that the attention in the very last timestep shifts towards the content, which aligns with our expectations.

\subsubsection{A Purely Content/Collaborative Approach}
Next, we try and understand how important our features are to the model. 

We train and test two variations, with and without the 128-dimensional content vectors, to see how important a content-based approach is for NERE. The impacts of these variations are demonstrated in Table \ref{comparison}.
\begin{table}[h!]
	\centering
\begin{tabular}{llll}
\hline
 & \textbf{Both} & \textbf{Content} & \textbf{Metadata} \\ \hline
\textbf{$R^2$} & 0.81 & 0.78 & 0.55 \\
\textbf{Recall@100} & 0.55 & 0.38 & 0.001 \\ \hline
\end{tabular}
\newline
\caption{Table \ref{comparison} demonstrates the importance of our content vectors.}
\label{comparison}
\end{table}

This shows that a hybrid (both collaborative and content-based) is clearly superior over either one independently. It is important to notice that a content-based approach will obtain a high $R^2$ score, since it is easy for the model to learn the underlying distribution, but will not recommend the appropriate set. This demonstrates the importance of various collaborative features that we explicitly include.

For example, the nearest neighbor for a set whose term and definition languages are in Spanish, is actually a set whose term and definition languages are in German. However, the model will continue to recommend sets with term and definition languages in German, since it has learned this from a user's prior history. This speaks to the importance of collaborative features in NERE.

On the whole, we have shown that NERE provides quality recommendations with which we can provide a deeply personalized experience for learning, and believe this results exceed expectations for our application.

\section{Conclusion \& Future Work}
In this work, we have proposed Neural Educational Recommendation Engine (NERE) to address the problem of personalized sequential recommendation in the Technology Enabled Learning (TEL) domain. By leveraging both content-based and collaborative features, our model can capture temporal trends in a user's history, and provide recommendations as to what they should learn next. By incorporating features such as attention and bidirectionality into our model, we were able to achieve a state of the art \textit{recall@100} score of 0.55. Moreover, we have performed an analysis of our model and have shown that it outperforms both a standalone content-based and collaborative approach. Lastly, we have shown that our model is learning from both the user and set metadata, in addition to content, by visualizing the attention mechanism. 

As to future work, we believe there is significant work left to be done in ranking the suggestions; there are significantly better ways to choose sets from a candidate pool than to recommend the $N$ closest neighbors. Furthermore, we believe that an attempt at canonicalizing similar sets would increase the \textit{Recall@100} metric, and should be explored. 
\section{Acknowledgements}
First and foremost, I would like to thank my mentors Dustin Stansbury and Shane Mooney for the exceptional support and mentorship throughout this project. Both of them were supportive, answered my many questions, and were quite open to letting me explore. Shane, thank you for providing much needed practical wisdom, for reviewing countless pull requests, and for providing much needed commentary on this paper. Dustin, thank you for the incredible knowledge about all things machine learning. This project wouldn't have been possible without you two.

I would also like to acknowledge Alex Pinchuk and Shaun Mitschrich for providing endless platform support throughout this project, including honoring my numerous requests for more compute.

Lastly, I would like to acknowledge the fabulous Quizlet team who provided incredible companionship throughout this summer, as well as my parents for supporting me throughout this process.

Keep on learning!
\bibliography{recommending_content}{}
\bibliographystyle{IEEEtran}
\end{document}